\documentclass[10pt]{iopart}
\usepackage{graphicx,hyperref}
\usepackage{multirow}
\usepackage[usenames, dvipsnames]{color}
\def\ket#1{|\,#1\,\rangle}
\def\Der#1#2{\frac{d  #1}{d #2}}
\def\bra#1{\langle\, #1\,|}
\def\braket#1#2{\langle\, #1\,|\,#2\,\rangle}
\def\proj#1#2{\ket{#1}\bra{#2}}
\def\mean#1{\langle \,#1\, \rangle}
\def\opone{\leavevmode\hbox{\small1\kern-3.8pt\normalsize1}}
\newcommand{\gguide}{{\it Preparing graphics for IOP Publishing journals}}
\def\abs#1{|\,#1\,|}
\usepackage{iopams}  
\begin{document}

\title[All-optical photon echo and  memory on a chip]{All-optical photon echo and memory on a chip}

\author{E.S.Moiseev$^{1,2}$, S.A. Moiseev$^{1*}$}
\address{$^1$ Kazan Quantum Center, Kazan National Research Technical University, 10 K. Marx, Kazan, 420111, Russia}
\address{$^2$ Department of Theoretical Physics, Institute of Physics, Kazan Federal University, 16a Kremlyovskaya St., Kazan, Russia, 420111}

\ead{s.a.moiseev@kazanqc.org$^{*}$}

\vspace{10pt}
\begin{indented}
\item[]November 2016
\end{indented}

\begin{abstract}
We demonstrate that a photon echo can be implemented by all-optical means using an array of on-chip high-finesse ring cavities whose parameters are chirped in such a way as to support equidistant spectra of cavity modes. When launched into such a system, a classical or quantum optical signal -- even a single-photon field -- becomes distributed between individual cavities, giving rise to prominent coherence echo revivals at well-defined delay times, controlled by the chirp of cavity parameters. This effect enables long storage times for high-throughput broadband optical delay and quantum memory.
\end{abstract}


\vspace{2pc}
\ioptwocol
\newcommand{\beq}{\begin{equation}}
\newcommand{\eeq}{\end{equation}}
\newcommand{\ba}{\begin{eqnarray}}
\newcommand{\ea}{\end{eqnarray}}
\newcommand{\bea}{\begin{eqnarray}}
\newcommand{\eea}{\end{eqnarray}}
\newcommand{\bma}{\begin{subequations}}
\newcommand{\ema}{\end{subequations}}
\newcommand{\bwt}{\begin{widetext}}
\newcommand{\ewt}{\end{widetext}}
\def\abs#1{\left\|\,#1\,\right\|}

\begin{abstract}
We demonstrate that a photon echo can be implemented by all-optical means using an array of on-chip high-finesse ring cavities whose parameters are chirped in such a way as to support equidistant spectra of cavity modes. When launched into such a system, a classical or quantum optical signal -- even a single-photon field -- becomes distributed between individual cavities, giving rise to prominent coherence echo revivals at well-defined delay times, controlled by the chirp of cavity parameters. This effect enables long storage times for high-throughput broadband optical delay and quantum memory.
\end{abstract}
A photon echo \cite{Kurnit1964,Allen1975} is a broad class of optical phenomena where a coherence induced in a quantum system by an optical field is emitted in a form of a well-resolved intense optical signal, similar to the spin echo in nuclear magnetic resonance. Over decades, photon echo has been in use as a powerful method of coherent spectroscopy, providing unique information on transient processes in gases, liquid, and solids \cite{Shen1984}. Photon-echo revivals in molecular rotational coherences have been shown to enable efficient quantum control of molecular alignment \cite{Stapelfeldt2003}, photochemical reactions \cite{Kartashov2012}, as well as synthesis of ultrashort field waveforms \cite{Fedotov2007}. In the era of quantum information, photon echo is viewed as a promising strategy for quantum data storage and quantum memories \cite{Moiseev2001, Riedmatten2008, Hedges2010, Hossein2011} (recent reviews see also in \cite{Lvov2009,Simon2010,Hammerer2010,Tittel2010}).

Here, we show that a photon echo can be implemented by purely optical means using an array of on-chip high-finesse ring cavities whose parameters are chirped in such a way as to support equidistant spectra of cavity modes. A classical or quantum optical signal launched into such a system becomes distributed between individual cavities, giving rise to prominent coherence echo revivals at well-defined delay times, controlled by the chirp of cavity parameters. This effect enables long storage times for high-throughput broadband optical delay and quantum memory.

\begin{figure}
\centering
\parbox{7cm}{
\includegraphics[width=7.5cm]{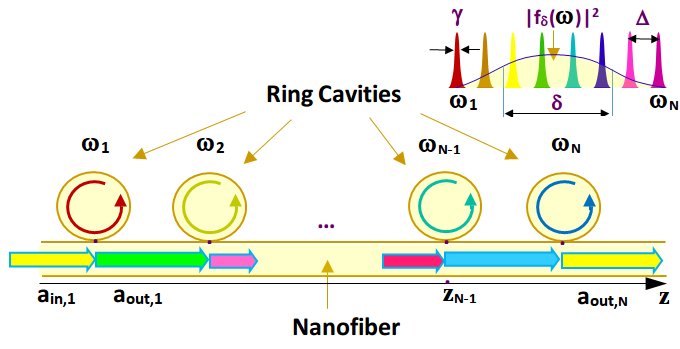}
\caption{An array of chirped ring cavities coupled to a common optical fiber. The spectrum of cavity modes $\omega_n=\omega_o+n\Delta$ with $\gamma<\Delta$ is shown in the inset. The spectrum of the input signal with a bandwidth $\delta$ is shown by the solid blue line.}
\label{scheme}}
\qquad
\begin{minipage}{7cm}
\includegraphics[width=7cm]{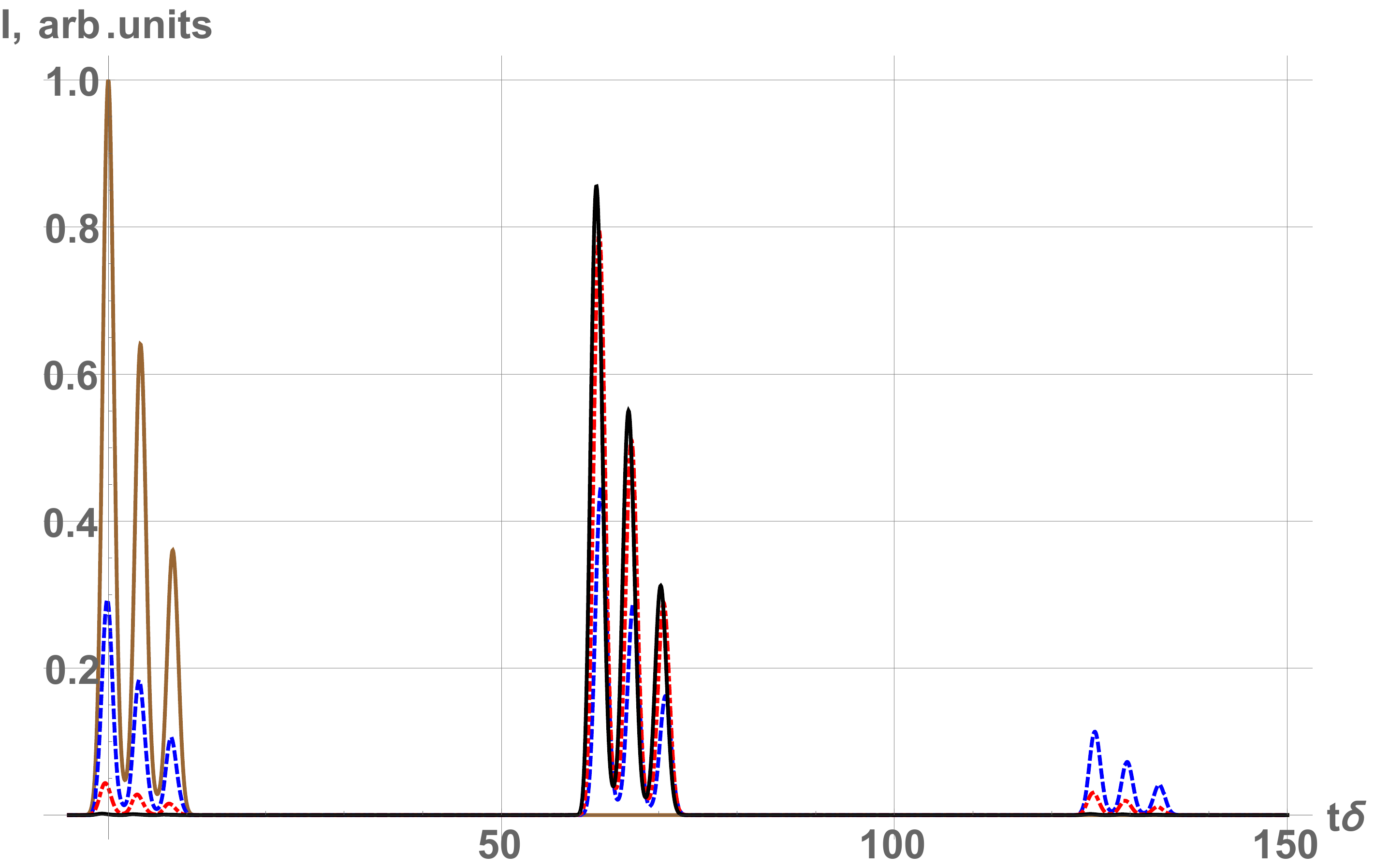}
\caption{The input and output field intensities (I=$|A_{in}(t)|^2$, $|A_{out}(t)|^2$) as functions of time (in units of $\delta^{-1}$) for a three-pulse input field (left solid brown curve) for $\kappa /\delta = 0.01$  (blue dashed curve), $0.025$ (red dash--dotted line), and $0.05$ (black line). The mode spacing is $\Delta = 0.1\delta$. The finesse of ring cavities is $F=\Delta/2\gamma =50$, number of cavities $N=61$.}
\label{MMode}
\end{minipage}

\begin{minipage}{7cm}
\includegraphics[width=7.5cm]{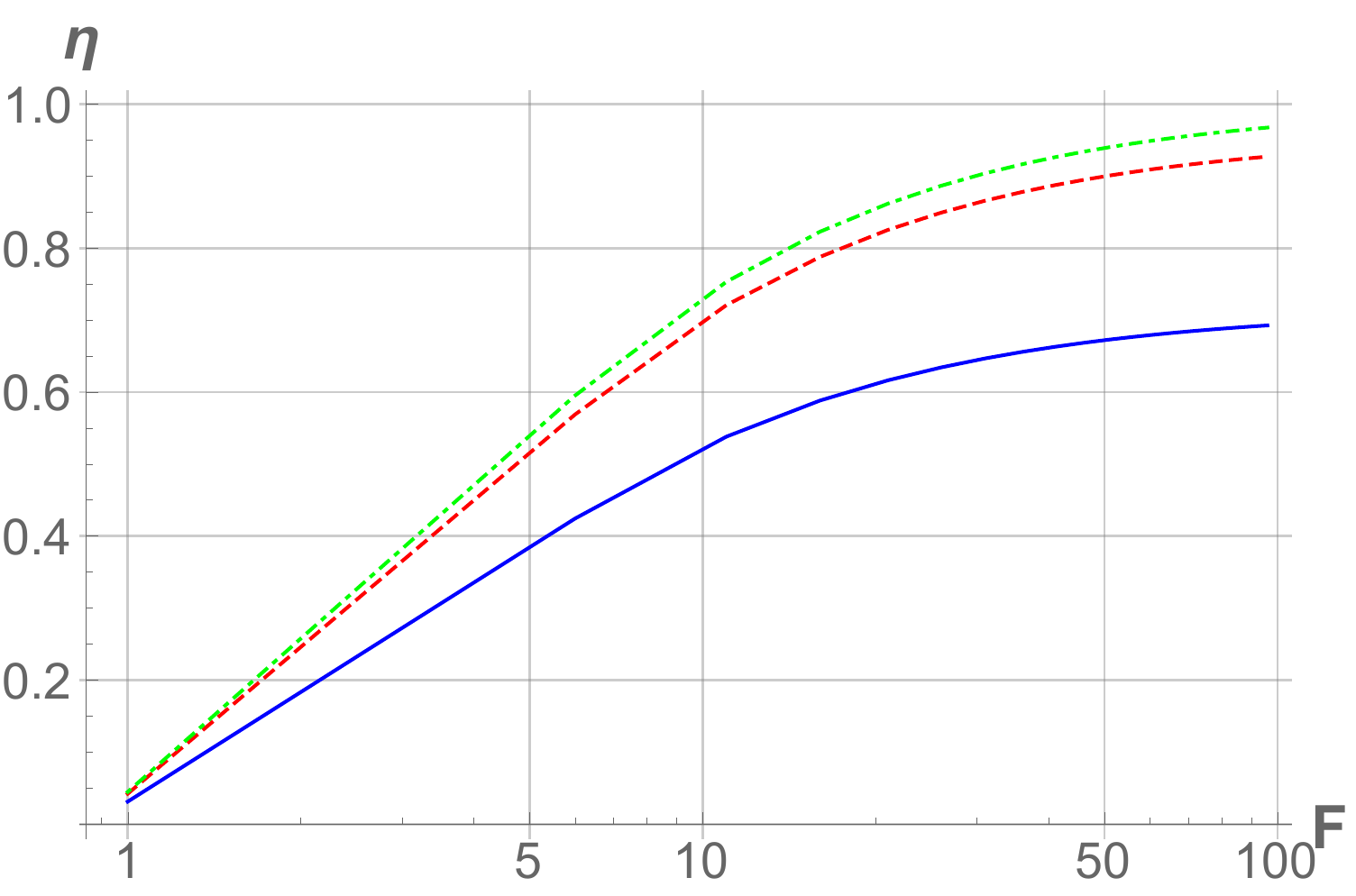}
\caption{Quantum efficiency $\eta$ of signal retrieval as a function of the cavity finesse $F=\frac{\Delta}{2\gamma}$ for a Gaussian pulse with $\delta=10\Delta$ in a cavity array with $N=61$ and $\kappa /\Delta = 0.10$ (blue solid line), $0.25$ (red dashed line), and $0.50$ (green dash--dotted line).}
\label{eta}
\end{minipage}
\end{figure}

We consider an array of $N$ single-mode high-finesse chirped cavities with an equidistant spectrum of modes with a mode spacing  $\Delta$ (Fig. \ref{scheme}). 
An optical field coupled into such an array remains distributed between the cavities until all the cavity modes can re-emit in phase, giving rise to an intense photon-echo signals at the output. 
With appropriate coupling between the nanofiber and the cavities, which is possible, e.g., with a fiber tapered to a submicron diameter \cite{Braginsky1999,Gorodetsky1999, Heebner2008}, the entire field stored in the cavity array can be retrieved within the first echo signal with a time delay $ t_{echo}\cong 2\pi/\Delta$.
We found the condition when this scheme can work as an efficient QM and classical time delay line  then we discuss possible experimental implementations and its further development for long lived storage.

In the considered chirped ring-cavity  (CRC) scheme (or frequency comb all-pass filter FC-APF), the spectrum of cavity modes is equidistant and consists of narrow lines centered at cavity eigenfrequencies $\omega_n$:
$\omega_n = \omega_0 + n \Delta-i\gamma_n$, where $\gamma_n<\Delta$ is the $n$th mode linewidth, and $n=0,1,...,N-1$.


Hamiltonian of the considered CRC system is written as $\hat{H} = \hat{H}_c + \hat{H}_f + \hat{V}$. $\hat{H}$, where $\hat{H}_c = \hbar \sum_{n=1}^N  \omega_n \hat{b}^{\dagger}_n \hat{b}_n$ is the Hamiltonian of $N$ ring cavities, $\hat{H}_f=\hbar\int d\omega\omega\hat{a}^{\dagger}_{\omega}{a}_{\omega}$ is the Hamiltonian of optical-fiber modes, $\hat{V}=\hbar \sum_n\int ̂d\omega (g_n\hat{b}^{\dagger}_n{a}_{\omega}e^{ik_{\omega}z_n}+g_n^*\hat{b}_n{a}^{\dagger}_{\omega}e^{-ik_{\omega}z_n})$ is the cavity--fiber interaction Hamiltonian, with $z_n$ being the coordinate of the $n$th cavity ($z_n\neq z_{m}$ for $n\neq m$) and $\hat{a}_{\omega}$,  $\hat{a}^{\dagger}_{\omega}$, $\hat{b}^{\dagger}_n$, and $\hat{b}_n$  being the Bose annihilation and creation operators of the optical-fiber and cavity modes.

An input optical light field is launched into the CRC system at the time moment $t=0$ along $z$ axis in Fig. \ref{scheme}.
Applying the input--output formalism \cite{Heebner2008,WallsMilburn2007} for the interaction of light field with ring cavity modes, which are assumed to be in the ground state at room temperature, we arrive at the following set of equations for the field amplitude inside the $n$th cavity: 
\bea{}
\frac{d\hat{b}_n}{dt} = -(i\omega_n +\frac{\kappa_n}{2}+\gamma_n)\hat{b}_n +\sqrt{\kappa_n} \hat{a}_{n,in}.
\eea{}

\noindent
To find the field amplitudes for each cavity  $\hat{b}_{n}$ and $\hat{a}_{in,n}$, we use the condition $ \hat{a}_{in,n} +\hat{a}_{out,n}=\sqrt{\kappa_n}\hat{b}_n$ \cite{WallsMilburn2007} and relate to the fiber fields  before and after interaction with the $n$th cavity through $\hat{a}_{in,n+1}=e^{-ik(z_{n+1}-z_{n})}\hat{a}_{out,n}$; $z_{n+1}>z>z_n$, where $\kappa_n$ is a coupling constant of the $n$-th cavity mode with the nanooptical fiber \cite{WallsMilburn2007}.  

With a Fourier transform $\hat{b}_n(t)=\int d\omega \hat{b}_n (\omega)e^{-i\omega t}$, $\hat{a}_{in,n}(t)=\int d\omega \hat{a}_{in,n} (\omega)e^{-i\omega t}$  in Eq.(1), we get

\bea{}
\hat{a}_{out,n}(\omega) = -\frac{\frac{1}{2}\kappa_n-\gamma_n-i(\omega_n-\omega)}{\frac{1}{2}\kappa_n+\gamma_n+i(\omega_n-\omega)}\hat{a}_{in,n}(\omega).
\label{aout}
\eea{}

\noindent
Since $\hat{a}_{n+1,in}=\hat{a}_{n,out}e^{ik_{\omega}(z_{n+1}-z_n)}$,
we find  

\bea{}
\hat{a}_{out,N}(\omega,z>z_{N}) = e^{i k_{\omega}(z-z_N)}U_{\omega}^N
\hat{a}_{in}(\omega), 
\label{aoutN}
\eea{}

\noindent
where  
\bea{}
U_{\omega}^m =(-1)^m e^{i k_{\omega}(z_m-z_1)} \prod_{n=1}^m\frac{\frac{1}{2}\kappa_n-\gamma_n-i(\omega_n-\omega)}{\frac{1}{2}\kappa_n+\gamma_n+i(\omega_n-\omega)},
\label{product}
\eea{}

\noindent
determines the fiber mode amplitude behind the $m$th cavity, $\hat{a}_{in}(\omega)=\hat{a}_{in,1}(\omega,z_1)$ is the Fourier transform of the input field, where $n$th cavity mode amplitude is

\bea{}
\hat b_n (t) = \int d\omega\beta_n(\omega)U_{\omega}^{(n-1)}e^{-i\omega t}\hat{a}_{in}(\omega), \label{cavityn}
\eea{}
\noindent
with $\beta_n(\omega)=\frac{\sqrt{\kappa_n}}{\frac{1}{2}\kappa_n+\gamma_n+i(\omega_n-\omega)}$.

As can be seen from Eqs.(\ref{aoutN}), (\ref{product}), the output field is independent of the order in which the cavities are arranged in the array. 
For a single-photon initial state $\ket{\psi_{\delta}}=\int d\omega f_{\delta} (\omega)\hat{a}_{in}^{\dagger}(\omega)\ket{\emptyset}$, where $\ket{\emptyset}$ is the vacuum state and $f_{\delta} (\omega)$ is the wave function normalized by $\int d\omega |f_{\delta}(\omega)|^2=1$ with a bandwidth $\delta$, the field amplitude in the $n$th cavity $B_n (t)=\bra{\emptyset}\hat b_n (t)\ket{\psi_{\delta}}$, the probability to find a photon in the $n$th cavity $P_n=\bra{\psi_{\delta}}\hat b^{\dagger}_n\hat b_n\ket{\psi_{\delta}}$, and the probability amplitude of the output light field $A_{out} (t,z)=\bra{\emptyset}\hat{a}_{out,N}(t,z)\ket{\psi_{\delta}}$ are given by   

\bea{}
B_n (t)&=&\int d\omega e^{-i\omega t} \beta_n(\omega)U_{\omega}^{(n-1)}f_{\delta}(\omega),\\ 
P_n (t) &=& B^*_n (t) B_n (t),\\
A_{out} (t,z) &=& \int d\omega e^{-i(\omega t-k_{\omega}(z-z_N))}U_{\omega}^N f_{\delta}(\omega).
\label{Aout}
\eea{}
The quantum efficiency $\eta$ of field retrieval in the first echo pulse is defined as
 
\bea{}
\eta (t) =\frac{\int dz A_{out}^{*}(t,z) A_{out}(t,z)}{\int dz A_{in}^{*}(t_0,z) A_{in}(t_0,z)},
\eea{}

\noindent
where $A_{in} (t_o,z)=\bra{\emptyset}\hat{a}_{in}(t_0,z)\ket{\psi_{\delta}}$ is the input field amplitude.

In Fig. \ref{MMode}, we plot the output field amplitude $A_{out} (t)$ calculated for a three-pulse input signal (left solid brown curve) as a function of time for three different values of the coupling constant $\kappa$ and fixed decay rate $\gamma$, with $\kappa=\kappa_1=...\kappa_N$, $\gamma=\gamma_1=...\gamma_N$, and the finesse and the cavity quality factor defined as $F=\Delta/(2\gamma)$ and $Q=\omega_0/(2\gamma)$. 
In the regime of weak coupling, $\kappa\ll\Delta$, some fraction of input light is transmitted through the entire CRC system, experiencing no delay, giving rise to a signal at $t\approx 0$ in Fig. \ref{MMode}. 
The remainder of the input field is distributed between the first, second, and third echo signals, observed at $t_{echo}$, $t_{echo,2}\approx2t_{echo}= 4\pi/\Delta$, and $t_{echo,3}\approx3t_{echo}$, respectively.    

\begin{figure}
\centering
\parbox{7cm}{
\includegraphics[width=7cm]{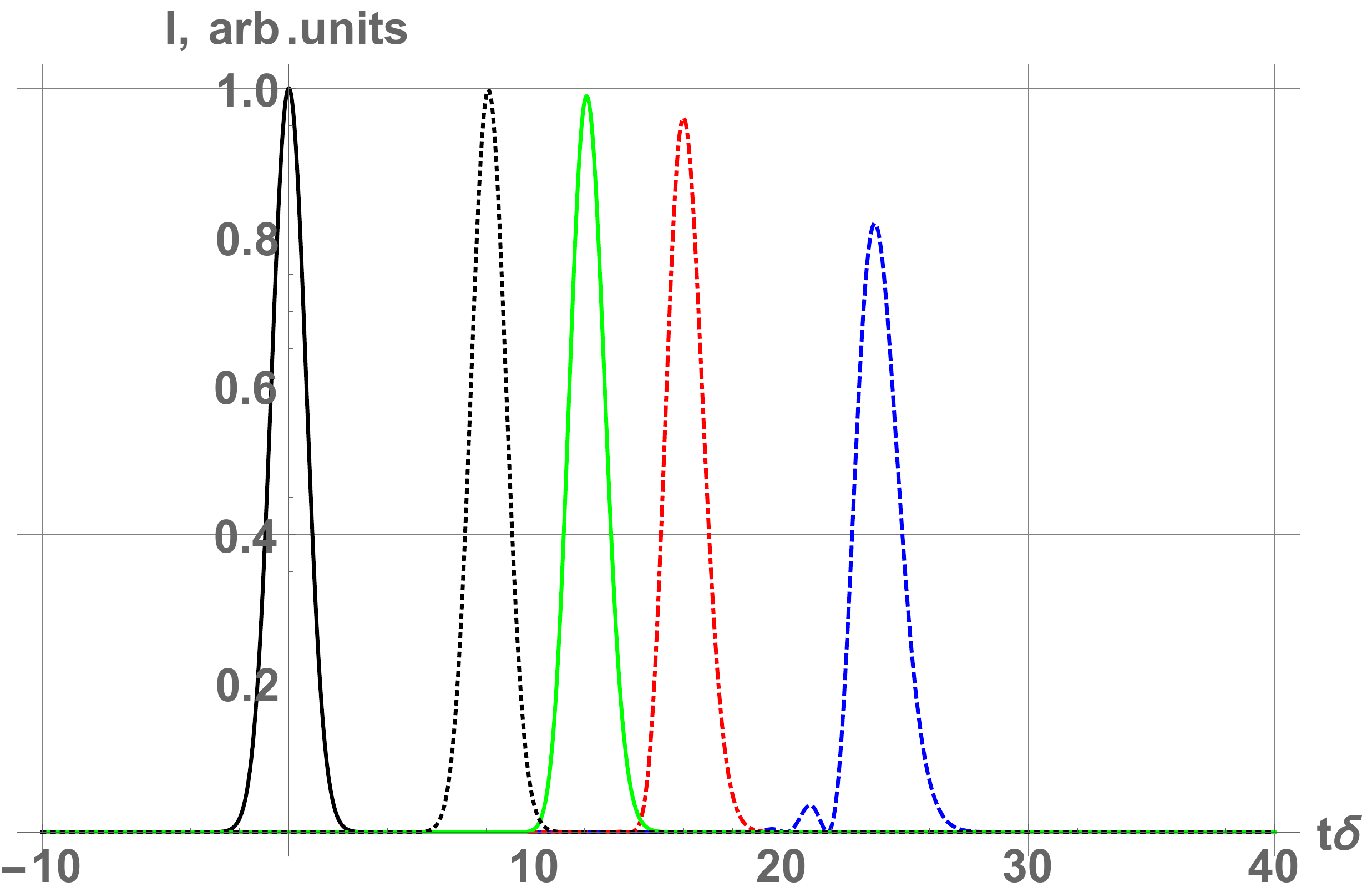}
\caption{ Intensities of the input and output light pulses (I=$|A_{in}(t)|^2$, $|A_{out}(t)|^2$) in the SCISSOR scheme  \cite{Heebner2002a} for  different coupling constants normalized to the spectral width of signal pulse $\kappa/\delta$: 5 (blue dashed), 7.5 (red dotdashed), 10 (green solid), 15 (black dotted); number of cavities $N=61$,  all the cavities have equal frequencies (i.e. $\Delta=0$), $\gamma=10^{-4}\delta$, $\delta$ is a spectral width of the input light pulse. The presented time delay is totally due to the interaction between the cavities and the incoming pulse. }
\label{SCISSOR}}

\begin{minipage}{7cm}
\includegraphics[width=7.cm]{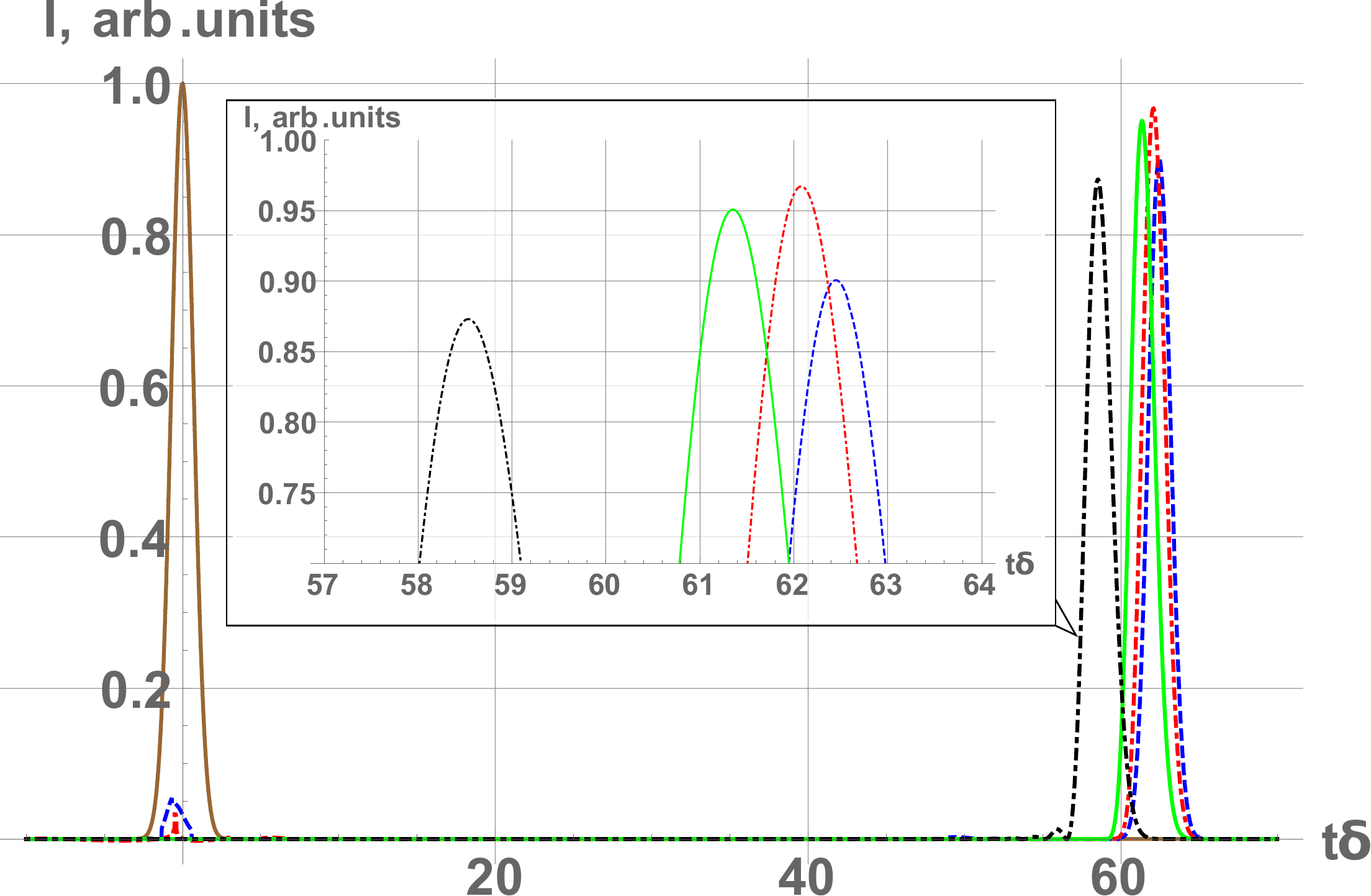}
\caption{Intensities of the input signal pulse and retrieved echo pulses (I=$|A_{in}(t)|^2$, $|A_{out}(t)|^2$) for ring cavity QM  vs coupling constant $\kappa$: $0.35\Delta$ (blue dashed), $0.5\Delta$ (red dotdashed), $1\Delta$ (green solid), $3\Delta$ (black dotted); spectral distance $\Delta=0.1 \delta$, number of cavities $N=61$, $\gamma=10^{-4}\delta$ (finesse $F=\Delta/(2\gamma)=500$), $\delta$ is a spectral width of the input light pulse. The calculated delay is totally due to the interaction between the cavities and incoming light pulse. }
\label{MMode2}
\end{minipage}
\end{figure}

Calculations presented in Fig. 3 show that, with $F >50$, the first echo provides signal retrieval with an efficiency $\eta>0.9$. As can be seen in Figs. \ref{MMode}, \ref{eta}, \ref{MMode2}, coupling constants $\kappa$ close to $0.5\Delta$ are required for efficient  signal retrieval.
The storage time in the considered cavity array is mainly determined by the frequency spacing $\Delta$, but is also sensitive to the coupling constant $\kappa$ (Fig.\ref{MMode2}). 
  
We emphasize that, while the CRC array considered here stores light -- classical or quantum -- in the form of a field distributed between individual cavities, its ability to provide long storage times is due to periodic coherence revivals, occurring at the instants of time when the fields circulating in individual cavities are all emitted in phase. 
In this respect, it is instructive to compare the delay-line performance of the CRC-array scheme considered here with a delay-line architecture based on the well-known all-pass filter (APF)  called also as a  side coupled integrated spaced sequence of resonators (SCISSOR) \cite{Heebner2002a,Heebner2002b,Xia2007}. 
With respect to CRC-array, SCISSOR scheme is characterized by the ring cavities with equal frequencies.
It is worth noting that SCISSOR together with coupled resonator optical waveguide (CROW) configuration  \cite{Yariv1999} are two  complementary basic standards of an on-chip integrated optical delay line (see \cite{Chremmos2010,Morichetti2012, Bogaerts2012}).
There are some common characteristics,  advantages and specific limitations in these two schemes.
Time delay in both the schemes is mainly determined by its bandwidth window where the effective group velocity of light can be reduced to $c/100$.  
Herein, CROW where the ring cavities are coupled with each other without additional common optical fiber can work within  wider spectral range even in a presence of weak strength of the connection between the resonators (i.e.  $\delta\gg\kappa$). 
In addition, although SCISSOR scheme can demonstrate a longer time delay but it is characterized by relatively higher losses that is caused by the significant role of resonant interaction with light field. 
Below we show that CRC-array scheme can considerably exceed the basic characteristics of SCISSOR scheme and to get the properties closer to CROW scheme.

A comparison of CRC and SCISSOR schemes shows (see Figs. \ref{SCISSOR}, \ref{MMode2} and Tables 1, 2) that both the   schemes can provide comparable maximum efficiencies, $\eta \approx 95$\%  for the same input light pulses. 
Moreover, both schemes can provide a superb fidelity, with the shape of the retrieved signal very accurately following the shape of the input pulse (Figs. 2, 4, 5). However, for comparable $\eta \approx 95$\% values, the delay times provided by a CRC array are substantially longer than the delay times attainable with the SCISSOR design. 
Finally, the coupling constants needed to achieve comparable delay times are much lower for CRC arrays. 
Specifically, calculations presented in Figs. 4 and 5 give $\tau\approx2\pi/\Delta\approx 62.8 /\delta$ for an CRC array with $\kappa=0.05\delta$ versus $\tau\approx 16.8/\delta$ for SCISSOR with $\kappa=7.5\delta$.  
Large difference in the coupling constants $\kappa$ becomes clear since the maximum bandwidth $\delta$  of SCISSOR is determined by the coupling constant $\simeq\kappa$ (or loaded Q-factor) of an individual resonator, when the total group delay results from a summation of the delays of all individual resonators.
In turn, CRC can provide larger time delay for broadband light pulses ($t\approx2\pi/\Delta\gg\delta^{-1}$ due to  the presence of spectral frequency combs $N\Delta\geq \delta$ for relatively weak coupling $\kappa\ll\delta$.  
Accordingly the predominant interaction of signal fields with CRC array  occurs at the off-resonant condition with the ring cavity modes that will provide lower losses similar to CROW.  

The proposed purely optical scheme  can be  experimentally implemented on  whispering gallery modes of the ring cavities \cite{Braginsky1999} and more easily using on-chip approach \cite{Lenz2001}. 
The cavity modes in these techniques  \cite{Braginsky1999} and in \cite{Vahala2003} are  characterized by very high intrinsic quality factors $Q=10^{8}$ and $Q=10^{10}$ where the  cavities can be efficiently coupled to the nanofiber \cite{Herr2014}. 
Below we analyze the possible implementation of CRC by means of on-chip array of ring  cavities \cite{Vahala2003}. 

Following the work \cite{Vahala2003}, we have the  parameters of single mode ring cavity: $\omega_0=1.26\cdot 10^{15} rad/sec$ corresponding to the telecommunication wavelength $
\lambda= 1.5\mu m$, cavity diameter $D = 90$ $\mu m$, free spectral range $\omega_{FSR}=4.96\cdot 10^{12} rad/sec$ and  $\gamma_n=\gamma=1/(2\tau_{crit})=10^{7} rad/sec $ determined by the intrinsic quality factor $Q=1.25\cdot 10^8$. 
By taking into account that total number of the cavities can reach $100$  \cite{Chremmos2010}, our consideration corresponds to an intermediate number $N=61$ that determines a maximum intermode spacing  $\Delta=\omega_{FSR}/60=8.27\cdot 10^{10} rad/sec$. 
The optimal coupling constant $\kappa=0.5\Delta=4.135\cdot 10^{10} rad/sec$ (see Fig.\ref{MMode2})  can be obtained experimentally at the appropriate spatial distance between the micro cavities and nanooptical waveguide \cite{Chremmos2010}.
For instance  $n$th cavity should be tuned to the   spectral detuning $n\Delta$  by using the matched spatial diameter $D_n\cong D-n\delta D$ (where $\delta D=\Delta D/\omega_0= 7.37\cdot 10^{-4} D=63$ $nm$). 
Size $63$ $nm$ is a large enough value for present nanooptical technologies  \cite{Heebner2008}.  
An arbitrary spectral offsets $n\Delta$ can be also operated in tunable microresonators rely on thermo-optics, or electro-optics and free-carrier dispersion control to vary the index of refraction \cite{Chremmos2010, Melloni2010}. 

In evaluation of efficient light pulse retrieval we apply dimensionless numerical results presented in Figs. \ref{SCISSOR}, \ref{MMode2} for different intermode spacings depending on the temporal duration of pulse as $\Delta=(10t_s)^{-1}$. 
That is, to demonstrate we only use the intermode spacing $\Delta=0.1\delta$ while a spectral width of light pulse $\delta=t_s^{-1}$ can also be in the range from few $\Delta$ to several dozen of $\Delta$.  
By using two possible intrinsic quality factors ($Q=10^8$,$Q=10^{10}$), we find the time delays $t_{delay}$ for SCISSOR and CRC schemes which correspond to the same quantum efficiency $\eta\approx 0.95$ (where $F=500$). 
For more evident representation, these data are also summarized in Table \ref{table:parameter} . 
The time delays in Table \ref{table:parameter} demonstrate a possibility for an efficient quantum storage of pico- and nanosecond pulses  where time delays in CRC scheme can reach $t_{echo}\cong 2$ ns for $t_s=32ps$,  $Q=10^8$ and $t_{echo}\cong 0.2\mu s$ for $t_s=3.2ns$, $Q=10^{10}$.

\begin{table}
\centering
\begin{tabular}{|l|c|c|r|}
\hline
{Q} & $t_{s}$ &{$t_{delay}$, SCISSOR} & {$t_{delay}$, CRC} \\ \hline
 \multirow{2}{*}{$10^8$} &  1.2 ps & 20.2 ps& 75.4 ps \\ \cline{2-4} 
&  32 ps&  0.54 ns& 2 ns \\ \hline 
\multirow{2}{*}{$10^{10}$} & 0.12 ns& 2.02 ns & 7.54 ns \\ 
\cline{2-4}  
& 3.2 ns& 54 ns & 0.2 $\mu$s \\\hline
\end{tabular}
\caption{\label{table:parameter} Time delay of the retrieved pulse in SCISSOR- and CRC schemes with quantum efficiency $\eta\approx 0.95$ related to the results of Figs. \ref{SCISSOR}, \ref{MMode2} for two intrinsic quality factors $Q$; coupling constant $\kappa =7.5 \delta$ for SCISSOR and $\kappa =0.05 \delta$ for CRC scheme, $\delta=t_s^{-1}$ is a spectral width of the signal pulse, $t_s$ - its temporal duration, $\delta=10\Delta$, number of microcavities N=61, finesse $F=500$.}
\end{table}

















Spatial size of the described array of $61$ cavities can be estimated as  $L=60 \cdot 100$ $\mu m$ $=6mm$ where  $100$ $\mu m$ is a distance between the centers of two nearest cavities. 
Additional time delay of light pulse due to the propagation in the common optical fiber $6$ $mm$ will be $\approx 3\cdot 10^{-2}$ nanoseconds (with fiber refractive index $1.5$) this time delay is shorter more than $60$ and $6\cdot 10^{3}$ times in comparison with $t_{delay}$ in CRC scheme for the quality factors $Q=10^8$ and $Q=10^{10}$ that indicates significant advantage of CRC for using in on-chip nanooptical schemes in comparison with an usual SCISSOR scheme.

Storage (delay) time in CRC scheme $t_{echo}\cong 2\pi/\Delta$ can be further increased for the cavities with smaller frequency spacing $\Delta$ (i.e. for lower finesse F) but at the expense of reducing the quantum efficiency. 
By using the numerical data presented in Fig.\ref{eta} , in Table \ref{table:parameter2} we present possible time delays for CRC scheme with appropriate quantum efficiencies.
Here, it is seen that one can get the time delay $t_{delay}\cong 334.7 ns$ for light pulse $t_s= 5.33 ns$  with quantum efficiency $\eta \approx 0.35$ (where $Q=10^8$). 
For higher quality factor $Q=10^{10}$ the time delay can reach $33.5 \mu s$. 
This time delay is shorter in one order of magnitude in comparison with the light lifetime   in a single ring cavity.   



\begin{table}
\centering
\begin{tabular}{|l|c|c|c|r|}
\hline
{Q(CRC)} & $F$ & $\eta$ &{$t_{s}$} & {$t_{delay}$} \\ \hline
 \multirow{2}{*}{$10^8$} & 10 &0.72 & 1.6 ns& 100.5 ns \\ \cline{2-5} 
& 3 & 0.35 & 5.33 ns& 334.7 ns \\ \hline 
\multirow{2}{*}{$10^{10}$} & 10& 0.72 & 0.16 $\mu$s & 10.05 $\mu$s \\ 
\cline{2-5}  
& 3& 0.35 & 0.533 $\mu$s & 33.5 $\mu$s \\\hline
\end{tabular}
\caption{\label{table:parameter2} Time delay of the retrieved pulse in CRC schemes in accordance with results of Figs. \ref{eta} for two different  finesse: $F=3$ $F=10$ and for two quality factors $Q$; coupling constant $\kappa =0.05\delta$, $\delta=10\Delta$ ($\delta=t_s^{-1}$), N=61.}
\end{table}


It is worth noting new nanooptical technologies which could be useful for fabrication of the studied CRC scheme. 
For example, it may be monolithic semiconductor microcavities and cavity arrays in photonic crystals and some others  \cite{Heebner2008,Chremmos2010}. 

Thus we  have shown that a system of frequency comb ring cavities one side coupled to the nanofiber can be used for implementation of on-chip broadband time delay line  operating at room temperature. 
In comparison with usual SCISSOR configuration 
\cite{Heebner2002a,Heebner2002b,Xia2007},  the proposed CRC scheme provides longer time delay with much more weaker coupling constant (with two orders of magnitudes) between the cavities and optical fiber that promises lower losses in practical usage.
These properties of CRC scheme essentially improves SCISSOR configuration and provides operation in a wider spectral range $\delta\gg \kappa$.

CRC scheme can be developed for quantum storage.
Comparing with free space photon echo AFC  protocol \cite{Riedmatten2008} promising for realization of optical quantum repeater, the described CRC scheme uses purely optical tools and can operate at room temperature.
It is also important that CRC  scheme provides an efficient light retrieval in forward direction that dramatically   facilitates the practical implementation of high quantum efficiency.
At the same time, a conventional AFC-protocol works efficiently only for backward light retrieval that however  requires the use of additional control laser pulses. 
Another non-obvious property of CRC scheme is an efficient quantum storage at the optimal coupling $\kappa=0.5\Delta$ that differs this scheme from the free space AFC protocol \cite{Riedmatten2008} and with its recent spatial-frequency version \cite{Tian2013}. 
The optimal coupling makes CRC protocol is  similar to the impedance matching QMs \cite{Afzelius2010,Moiseev2010,Moiseev2013} although CRC does not use any common single mode resonator providing an optimal coupling with all the CRC modes. 
This property exposes another physical nature in the emergence of optimal conditions in the CRC scheme and as in recent work on quantum RAM \cite{EMoiseev2016} again indicating the possibility of more common manifestation of the impedance matching effects for light field evolution  in the integrated optical devices.
 
On demand retrieval in CRC scheme can be realized by adiabatically fast equalizing the cavity frequencies after the complete absorption of an input light pulse but before its irradiation in echo pulse at $t\cong 2\pi/\Delta$. 
The frequency controlling  the ring cavity modes can be implemented by using existed and intensively elaborated methods \cite{Chremmos2010, Melloni2010, Canciamilla2010} providing sufficiently fast  switching the microresonator frequencies up to ten and hundreds $GHz$ bandwidth.
The equalized cavity frequencies will lead to freezing  relative phases of CRC modes and to complete stopping the input light pulse in the CRC arrays, respectively.
Further light retrieval will be possible only after recovering the initial cavity frequencies at the subsequent rephasing of the excited cavity modes.
Although the maximum storage time in this case will be limited by the intrinsic cavity mode Q-factor and by additional losses caused by the non ideal switching but this technique opens a direct way for the on demand fast retrieval of the stored light pulses.

More long-lived storage in CRC array can be implemented by incorporating resonant atoms in the ring cavity volume. 
Herein, one can reversibly transfer the cavity mode excitation to the long-lived  atomic/spin coherence (for example on the hyperfine sublevels of a NV-center \cite{Balasubramanian, Fuchs,  Yang,  Maurer, Tsukanov2013}) by applying an additional resonant laser field. 
In particular, for a single photon storage we can put only single atom (with optical $\lambda$ transitions) in each high quality ring cavity. 
By taking into account our studies, it means that only 61 such atoms  will be independently controlled for the storage of a single photon field on long-lived atomic transition \cite{10}. 
Since a single atom is characterized by narrower resonant transitions, the proposed scheme seems to be more realistic for use as an efficient broadband optical quantum memory and repeater operating at room temperature, for example, when using NV center as a long-lived three level atom. 
All the discussed observations indicates promising opportunities of CRC scheme for practical implementation of the frequency comb based quantum repeater.    
The detailed studies of long-lived CRC based quantum memory schemes will be a subject of our further works.  


The described properties of all-optical photon echo on the chirp ring cavity array coupled with nanofibers demonstrate a promising credit to the room temperature time delay line and optical quantum memory suitable for application in on-chip optical schemes, quantum processing and communication.   

Authors are grateful to A.M.Zheltikov for useful discussions  and M.L. Gorodetsky and T. Chaneliere for valuable comments, E.S.M and S.A.M. also thank the Russian Scientific Fund through for its financial support of this work through the grant no. 14-12-01333.


\begin{thebibliography}{99}
\bibitem{10} Ten cavities with single atoms are also possible for an efficient storage provided an appropriate choice of the coupling constant $\kappa$ and spectral detuning $\Delta$.
\end{thebibliography}
\end{document}